\documentclass[aps,prl,10pt,amsmath,twocolumn,floatfix,superscriptaddress,longbibliography]{revtex4-1}
\usepackage[dvipsnames]{xcolor}
\usepackage[colorlinks=true,bookmarks=false,linkcolor=NavyBlue,urlcolor=NavyBlue,citecolor=NavyBlue,breaklinks]{hyperref}
\usepackage{amsmath,amsfonts,amssymb,amsthm}
\usepackage{mathtools}
\usepackage{tabularx}
\usepackage{graphicx}
\usepackage[italicdiff]{physics}
\usepackage{times}

\newcommand{\avg}[1]{\langle#1\rangle}
\newcommand{\Avg}[1]{\left\langle#1\right\rangle}

\begin{document}

\title{A time-symmetric generalization of quantum mechanics}

\author{Mankei Tsang}
\email{mankei@nus.edu.sg}
\homepage{https://blog.nus.edu.sg/mankei/}
\affiliation{Department of Electrical and Computer Engineering,
  National University of Singapore, 4 Engineering Drive 3, Singapore
  117583}

\affiliation{Department of Physics, National University of Singapore,
  2 Science Drive 3, Singapore 117551}

\date{\today}

%\pacs{42.50.Wk, 03.65.Ta, 42.65.Yj}

\begin{abstract}
  I propose a time-symmetric generalization of quantum mechanics that
  is inspired by scattering theory. The model postulates two
  interacting quantum states, one traveling forward in time and one
  backward in time. The interaction is modeled by a unitary scattering
  operator. I show that this model is equivalent to pseudo-unitary
  quantum mechanics.
\end{abstract}

\maketitle
After a century of research, the principles of quantum mechanics
remain the same as those proposed by Schr\"odinger and Heisenberg: a
vector in a Hilbert space represents the state of a physical system,
and a unitary operator models the dynamics. Early attempts at relaxing
the unitarity requirement \cite{dirac42,pauli43} were soon abandoned
in favor of a theory of quantized fields that retains it
\cite{weinberg}. In view of the mired progress in unifying quantum
mechanics with general relativity, however, non-unitary
generalizations of quantum mechanics have received renewed interest in
recent years. One interesting idea that goes back to Dirac and Pauli
\cite{dirac42,pauli43} is to replace the unitary operator with a
pseudo-unitary operator that conserves a scalar product with an
indefinite metric \cite{mosta10,bender}.

Here I propose an alternative approach to generalizing quantum
mechanics that is arguably more intuitive and is inspired by
scattering theory. In scattering theory, one considers incoming and
outgoing waves that travel in the spatial dimensions and are coupled
through scatterers. As quantum mechanics is a generalization of wave
mechanics, it is tempting to apply scattering theory to quantum states
in the time dimension as well, in which case one needs to postulate
two states, one traveling forward in time and one backward in
time. Coupling the two states via a scattering operator leads to a new
time-evolution operator, which is given by the so-called Potapov
transform \cite{neretin,potapov55} of the scattering operator. If the
scattering operator is unitary, the time-evolution operator turns out
to be pseudo-unitary, thereby establishing a correspondence between
the time-symmetric model set forth and the pseudo-unitary model in the
literature.

With the interaction of the two states moving in opposite time
directions, the model permits time travel, which is, of course,
another interesting but controversial topic in physics.  Given the
model's equivalence with the pseudo-unitary model, the possibility of
time travel may explain why the latter can violate certain principles
of standard physics, such as the no-signaling and no-cloning laws
\cite{lee14,zhan20}.  One should not take the time travel and the
violation of conventional principles as a failure of the models,
however, as the models allow time travel only in a rigid mechanistic
manner and there is no reason to believe that those conventional
principles are fundamental and can survive new physics.

I stress that the mathematics here is elementary and well established
in scattering theory, transmission-line theory, and optics when it
comes to waves traveling in the spatial dimensions; see, in
particular, the seminal works of Potapov \cite{potapov55} and
Redheffer \cite{redheffer62}. The transfer-matrix method in optics is
perhaps the simplest example of the scattering theory
\cite{born_wolf}. It is also known that the scattering problem with
the time-independent Schr\"odinger equation in one spatial dimension
has a (2-by-2) pseudo-unitary transfer matrix (see footnote on p.~33
of Ref.~\cite{mosta20}).  The key new insight of this paper is that
the formalism, though mathematically simple, offers a principled way
to generalize time evolution in quantum mechanics for Hilbert spaces
with arbitrary dimensions, putting time on a more equal footing with
space in this fundamental law of physics. I also note that the
approach here seems to share some conceptual similarities with the
time-symmetric treatment of classical electrodynamics \cite{wheeler49}
and the Dirac equation \cite{feynman49} by Wheeler and Feynman,
although their approach was later subsumed by the unitary quantum
field theory \cite{weinberg}.  Some works on quantum measurement
theory also consider the combination of two quantum states
\cite{watanabe,aharonov_rohrlich,cramer86,smooth}, but those merely
offer alternative interpretations or applications of standard quantum
mechanics and do not modify it.

\begin{figure}[htbp!]
\centerline{\includegraphics[width=0.3\textwidth]{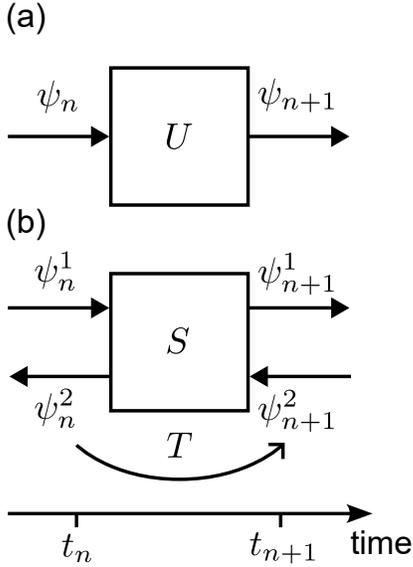}}
\caption{\label{block}(a) The standard quantum model in the
  Schr\"odinger picture, where the quantum state $\psi$ evolves
  forward in time from $\psi_n$ to $\psi_{n+1}$ via a unitary operator
  $U$. (b) The time-symmetric model, which involves two states
  $\psi^1$ and $\psi^2$ traveling forward and backward in time. Their
  interaction is modeled as an input-output relation in terms of a
  unitary scattering operator $S$, or equivalently a forward-time
  relation in terms of a pseudo-unitary transfer operator $T$.}
\end{figure}

To set the stage, I first review standard quantum mechanics. Let
$\mathcal H$ be a complex Hilbert space with an inner product denoted
by $\avg{u,v} \in \mathbb C$ with $u,v \in \mathcal H$.  Let
$\psi_n \in \mathcal H$ be a Hilbert-space vector that models the
quantum state of a physical system in the Schr\"odinger picture at
time $t_n$.  If the time evolution from $t_n$ to $t_{n+1}$ is modeled
by a linear operator $U$ on $\mathcal H$, such that
\begin{align}
\psi_{n+1} &= U \psi_n,
\end{align}
and the norm is required to be conserved in time, viz.,
\begin{align}
\Avg{\psi_{n+1},\psi_{n+1}} &= \Avg{\psi_n,\psi_n},
\end{align}
then $U$ must be unitary (I do not consider antilinear
operators). $\avg{\psi,\psi}$ is commonly regarded as the total
probability, although I do not prescribe any physical meaning to the
conservation law in the following to avoid premature interpretations.
Figure~\ref{block}(a) illustrates this standard quantum model by a
block diagram.

The proposed time-symmetric model assumes instead that there are two
worlds, one traveling forward in time and one backward in time. Let
$\mathcal H^1$ and $\mathcal H^2$ be the Hilbert spaces for the
forward and backward worlds, respectively, and let the total Hilbert
space for the two worlds be the direct sum
$\mathcal H^1 \oplus \mathcal H^2$. Let
\begin{align}
\psi_n = \psi_n^1 \oplus \psi_n^2 \in \mathcal H^1 \oplus \mathcal H^2
\end{align}
be the total state at time $t_n$.  Frame the problem as a scattering
problem, where $\psi_n^1$ and $\psi_{n+1}^2$ are the incoming waves
while $\psi_{n+1}^1$ and $\psi_n^2$ are the outgoing waves, as
illustrated by Figure~\ref{block}(b). If the scattering is linear, it
can be modeled by a scattering operator $S$ on
$\mathcal H^1\oplus\mathcal H^2$, such that
\begin{align}
\begin{pmatrix}\psi_{n+1}^1 \\ \psi_n^2\end{pmatrix}
&= \begin{pmatrix} S^{11} & S^{12} \\ S^{21} & S^{22} \end{pmatrix}
\begin{pmatrix}\psi_n^1 \\ \psi_{n+1}^2\end{pmatrix},
\label{scatter_matrix}
\end{align}
where $S$ is expressed as a matrix of four operators
$S^{jk}:\mathcal H^k \to \mathcal H^j$. In particular, the $S^{12}$
and $S^{21}$ operators model the interactions between the two
worlds. A reasonable conservation law that can be borrowed from
scattering theory is
\begin{align}
\Avg{\psi_{n+1}^1,\psi_{n+1}^1} + \Avg{\psi_n^2,\psi_n^2}
&= \Avg{\psi_n^1,\psi_n^1} + \Avg{\psi_{n+1}^2,\psi_{n+1}^2},
\end{align}
which implies that $S$ must be unitary. Note that this unitarity is
applied to the total dynamics of the two worlds and is different from
the one-world unitarity of standard quantum mechanics.

\begin{figure}[htbp!]
\centerline{\includegraphics[width=0.48\textwidth]{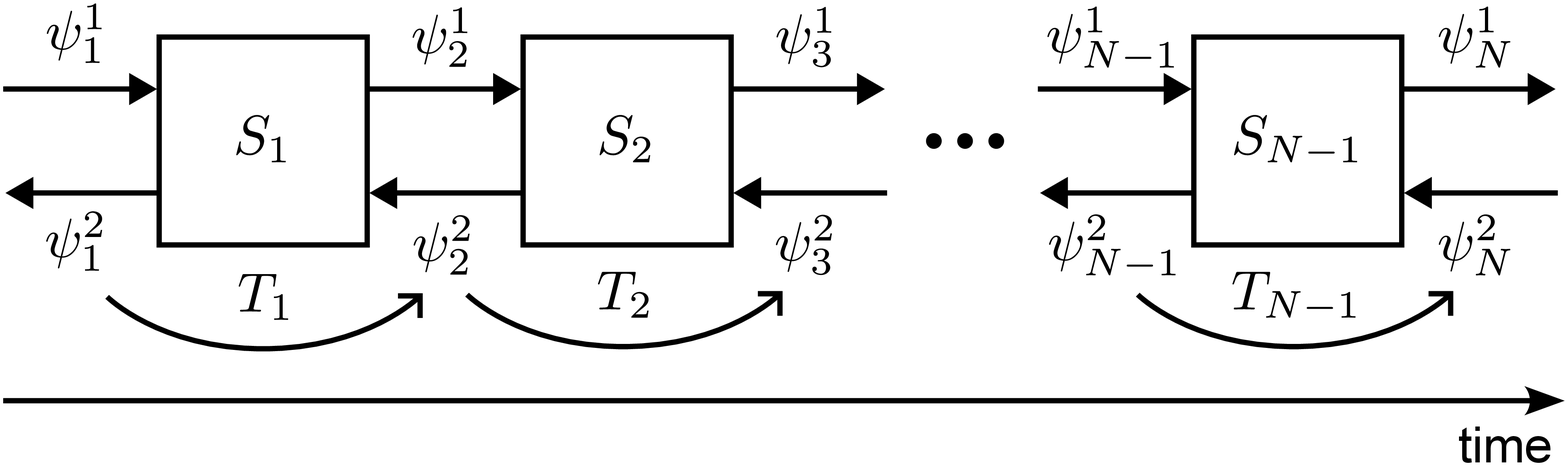}}
\caption{\label{cascade}A sequence of unitary interactions between the
  forward and backward states can be modeled by a sequence of
  pseudo-unitary transfer operators $T_1, T_2,\dots$}
\end{figure}

The conservation law can be rewritten as
\begin{align}
\Avg{\psi_{n+1}^1,\psi_{n+1}^1} -\Avg{\psi_{n+1}^2,\psi_{n+1}^2}
&= \Avg{\psi_n^1,\psi_n^1} -\Avg{\psi_n^2,\psi_n^2},
\end{align}
such that it can be interpreted as a conservation
of the single-time scalar product
\begin{align}
\Avg{\psi_{n+1},J\psi_{n+1}} &= \Avg{\psi_n,J\psi_n}
\end{align}
with the indefinite metric
\begin{align}
J &\equiv \begin{pmatrix} I^1 & 0\\ 0 & -I^2\end{pmatrix},
\end{align}
where $I^j$ is the identity operator on $\mathcal H^j$. Let $T$ be the
transfer operator on $\mathcal H^1\otimes\mathcal H^2$ that relates
$\psi_n$ at one time to $\psi_{n+1}$ at a forward time, viz.,
\begin{align}
\psi_{n+1} &= T\psi_n.
\end{align}
To be more explicit, 
\begin{align}
\begin{pmatrix}\psi_{n+1}^1 \\ \psi_{n+1}^2\end{pmatrix}
&= \begin{pmatrix} T^{11} & T^{12} \\ T^{21} & T^{22} \end{pmatrix}
\begin{pmatrix}\psi_n^1 \\ \psi_{n}^2\end{pmatrix},
\end{align}
which is similar to Eq.~(\ref{scatter_matrix}) in that $T$ is also
partitioned into four operators
$T^{jk}:\mathcal H^k \to \mathcal H^j$, although $\psi_n^2$ and
$\psi_{n+1}^2$ have switched places here. As $T$ is linear and
conserves the $J$-weighted scalar product, it must be pseudo-unitary,
or more precisely $J$-unitary, viz.,
\begin{align}
T^\dagger J T &= J,
\end{align}
where $\dagger$ denotes the adjoint in the usual sense. General
forward-time evolution is then described by a sequence of $J$-unitary
operators $T_1, T_2,\dots$ in the form
\begin{align}
\psi_{N} &= T_{N-1} \dots T_2 T_1\psi_1,
\end{align}
as depicted by Fig.~\ref{cascade}. This is precisely the model of
pseudo-unitary quantum mechanics \cite{mosta10}.

If $S^{22}$ is invertible, the transfer operator is related to the
scattering operator through the Potapov transform
\cite{neretin,potapov55}
\begin{align}
T &= \Pi(S) \equiv 
\begin{pmatrix} S^{11} -S^{12} (S^{22})^{-1} S^{21} 
& S^{12}(S^{22})^{-1} \\ -(S^{22})^{-1} S^{21} & (S^{22})^{-1} \end{pmatrix}.
\end{align}
The transform is straightforward to derive: start from
Eq.~(\ref{scatter_matrix}) and express $\psi_{n+1}^1$ and
$\psi_{n+1}^2$ in terms of $\psi_n^1$ and $\psi_{n}^2$, thus switching
the places of $\psi_n^2$ and $\psi_{n+1}^2$ in the matrix relation. It
is then obvious that applying the Potapov transform again to $T$
should give back the scattering operator, viz.,
\begin{align}
\Pi(T) &= \Pi[\Pi(S)] = S.
\end{align}
In other words, the pseudo-unitary model can be transformed back to
the unitary time-symmetric model, and the two models are equivalent,
as long as $T^{22}$ and $S^{22}$ are invertible. 

\begin{figure}[htbp!]
\centerline{\includegraphics[width=0.3\textwidth]{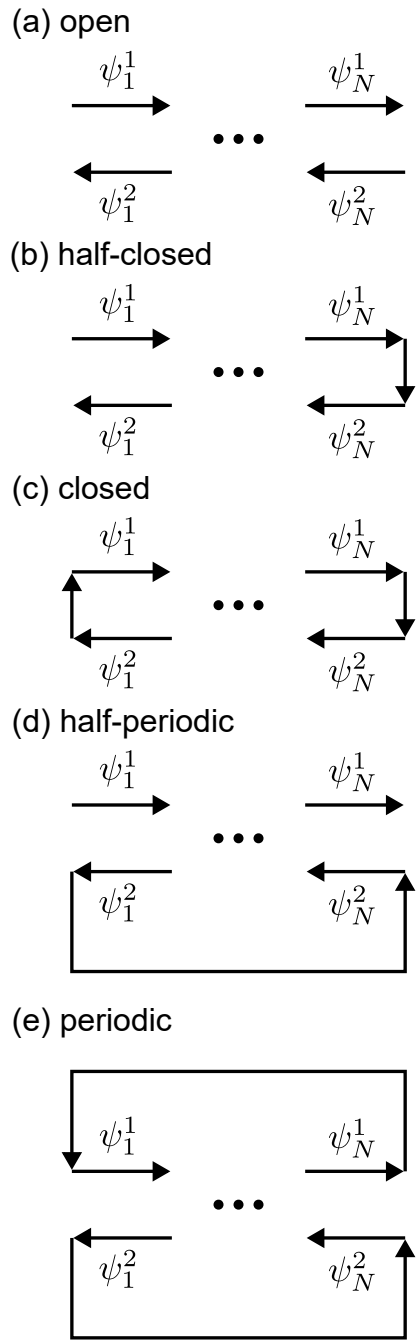}}
\caption{\label{boundaries}Different types of boundary conditions for
  the time-symmetric model.}
\end{figure}

To arrive at a definite solution, one also needs to specify the
boundary conditions at the initial time $t_1$ and the final time
$t_N$. In standard quantum mechanics, one may assume an initial state
$\psi_1$ or a final $\psi_N$ as a boundary condition, or use the
periodic boundary condition $\psi_1 = \psi_N$ to restrict the set of
solutions. With the two worlds in the time-symmetric model, there are
now more possible types of boundary conditions, as illustrated by
Fig.~\ref{boundaries}.  The first type is the open condition, which is
analogous to the usual scattering problem.  The inputs $\psi_1^1$ and
$\psi_N^2$ can be used as the boundary conditions, although one may
also pick any two states from
$\{\psi_1^1,\psi_N^2, \psi_N^1, \psi_1^2\}$ as long as the reverse
problem remains well posed. The second type, the half-closed
condition, involves a reflection that sets $\psi_N^2 = \psi_N^1$, and
only $\psi_1^1$ or $\psi_1^2$ remains to be specified. The third type,
where the system is closed at both ends, implies that the total world
behaves like a Fabry-P\'erot cavity, and a nonzero solution can exist
only as a superposition of its eigenmodes. The fourth type, which
introduces a periodic condition to the backward world, turns the
backward world into a ring cavity and restores unitarity to the
relation between the initial and final states of the forward world,
although the intermediate interaction between the two worlds still
makes the model different from the standard one-world model. The final
type, the fully periodic condition, is similar to the closed condition
but in a ring geometry.

Given the correspondence with scattering theory, the time-symmetric
model can be experimentally simulated by a photonic circuit with a
recirculating mesh \cite{bogaerts20} if one of the spatial dimensions
is used to represent time. It is well known that a forward-only mesh
can simulate a unitary system \cite{reck94}; a recirculating one
offers new possibilities in quantum simulation.

As a model of the universe, the time-symmetric model is consistent
with standard quantum mechanics and existing experimental evidence if
the two worlds interact so weakly that the transfer operator is close
to unitary and current experiments cannot detect the interaction. The
unitary operator of the standard model is expected to be an
approximation of $S^{11}$ and $T^{11}$ for the forward world; the next
questions are how one should model the rest of the operators and how
the deviation from standard unitarity may be tested by experiments.

%\pagebreak

\bibliography{research2}

\begin{thebibliography}{20}%
\makeatletter
\providecommand \@ifxundefined [1]{%
 \@ifx{#1\undefined}
}%
\providecommand \@ifnum [1]{%
 \ifnum #1\expandafter \@firstoftwo
 \else \expandafter \@secondoftwo
 \fi
}%
\providecommand \@ifx [1]{%
 \ifx #1\expandafter \@firstoftwo
 \else \expandafter \@secondoftwo
 \fi
}%
\providecommand \natexlab [1]{#1}%
\providecommand \enquote  [1]{``#1''}%
\providecommand \bibnamefont  [1]{#1}%
\providecommand \bibfnamefont [1]{#1}%
\providecommand \citenamefont [1]{#1}%
\providecommand \href@noop [0]{\@secondoftwo}%
\providecommand \href [0]{\begingroup \@sanitize@url \@href}%
\providecommand \@href[1]{\@@startlink{#1}\@@href}%
\providecommand \@@href[1]{\endgroup#1\@@endlink}%
\providecommand \@sanitize@url [0]{\catcode `\\12\catcode `\$12\catcode
  `\&12\catcode `\#12\catcode `\^12\catcode `\_12\catcode `\%12\relax}%
\providecommand \@@startlink[1]{}%
\providecommand \@@endlink[0]{}%
\providecommand \url  [0]{\begingroup\@sanitize@url \@url }%
\providecommand \@url [1]{\endgroup\@href {#1}{\urlprefix }}%
\providecommand \urlprefix  [0]{URL }%
\providecommand \Eprint [0]{\href }%
\providecommand \doibase [0]{http://dx.doi.org/}%
\providecommand \selectlanguage [0]{\@gobble}%
\providecommand \bibinfo  [0]{\@secondoftwo}%
\providecommand \bibfield  [0]{\@secondoftwo}%
\providecommand \translation [1]{[#1]}%
\providecommand \BibitemOpen [0]{}%
\providecommand \bibitemStop [0]{}%
\providecommand \bibitemNoStop [0]{.\EOS\space}%
\providecommand \EOS [0]{\spacefactor3000\relax}%
\providecommand \BibitemShut  [1]{\csname bibitem#1\endcsname}%
\let\auto@bib@innerbib\@empty
%</preamble>
\bibitem [{\citenamefont {Dirac}(1942)}]{dirac42}%
  \BibitemOpen
  \bibfield  {author} {\bibinfo {author} {\bibfnamefont {Paul Adrien~Maurice}\
  \bibnamefont {Dirac}},\ }\bibfield  {title} {\enquote {\bibinfo {title}
  {Bakerian {Lecture} - {The} physical interpretation of quantum mechanics},}\
  }\href {\doibase 10.1098/rspa.1942.0023} {\bibfield  {journal} {\bibinfo
  {journal} {Proceedings of the Royal Society of London. Series A. Mathematical
  and Physical Sciences}\ }\textbf {\bibinfo {volume} {180}},\ \bibinfo {pages}
  {1--40} (\bibinfo {year} {1942})}\BibitemShut {NoStop}%
\bibitem [{\citenamefont {Pauli}(1943)}]{pauli43}%
  \BibitemOpen
  \bibfield  {author} {\bibinfo {author} {\bibfnamefont {W.}~\bibnamefont
  {Pauli}},\ }\bibfield  {title} {\enquote {\bibinfo {title} {On {Dirac}'s
  {New} {Method} of {Field} {Quantization}},}\ }\href {\doibase
  10.1103/RevModPhys.15.175} {\bibfield  {journal} {\bibinfo  {journal}
  {Reviews of Modern Physics}\ }\textbf {\bibinfo {volume} {15}},\ \bibinfo
  {pages} {175--207} (\bibinfo {year} {1943})}\BibitemShut {NoStop}%
\bibitem [{\citenamefont {Weinberg}(1995)}]{weinberg}%
  \BibitemOpen
  \bibfield  {author} {\bibinfo {author} {\bibfnamefont {S.}~\bibnamefont
  {Weinberg}},\ }\href@noop {} {\emph {\bibinfo {title} {The Quantum Theory of
  Fields. Vol. I. Foundations}}}\ (\bibinfo  {publisher} {Cambridge University
  Press},\ \bibinfo {address} {Cambridge},\ \bibinfo {year} {1995})\BibitemShut
  {NoStop}%
\bibitem [{\citenamefont {Mostafazadeh}(2010)}]{mosta10}%
  \BibitemOpen
  \bibfield  {author} {\bibinfo {author} {\bibfnamefont {Ali}\ \bibnamefont
  {Mostafazadeh}},\ }\bibfield  {title} {\enquote {\bibinfo {title}
  {Pseudo-{H}ermitian representation of quantum mechanics},}\ }\href {\doibase
  10.1142/S0219887810004816} {\bibfield  {journal} {\bibinfo  {journal}
  {International Journal of Geometric Methods in Modern Physics}\ }\textbf
  {\bibinfo {volume} {07}},\ \bibinfo {pages} {1191--1306} (\bibinfo {year}
  {2010})}\BibitemShut {NoStop}%
\bibitem [{\citenamefont {Bender}(2019)}]{bender}%
  \BibitemOpen
  \bibfield  {author} {\bibinfo {author} {\bibfnamefont {Carl~M.}\ \bibnamefont
  {Bender}},\ }\href {\doibase 10.1142/q0178} {\emph {\bibinfo {title} {PT
  Symmetry in Quantum and Classical Physics}}}\ (\bibinfo  {publisher} {World
  Scientific},\ \bibinfo {address} {Singapore},\ \bibinfo {year}
  {2019})\BibitemShut {NoStop}%
\bibitem [{\citenamefont {Neretin}(2011)}]{neretin}%
  \BibitemOpen
  \bibfield  {author} {\bibinfo {author} {\bibfnamefont {Yurii~A.}\
  \bibnamefont {Neretin}},\ }\href
  {https://www.ems-ph.org/books/book.php?proj_nr=126} {\emph {\bibinfo {title}
  {{Lectures on Gaussian Integral Operators and Classical Groups}}}}\ (\bibinfo
   {publisher} {European Mathematical Society},\ \bibinfo {address} {Zurich},\
  \bibinfo {year} {2011})\BibitemShut {NoStop}%
\bibitem [{\citenamefont {Potapov}(1955)}]{potapov55}%
  \BibitemOpen
  \bibfield  {author} {\bibinfo {author} {\bibfnamefont {V.~P.}\ \bibnamefont
  {Potapov}},\ }\bibfield  {title} {\enquote {\bibinfo {title} {The
  multiplicative structure of {J}-contractive matrix functions},}\ }\href@noop
  {} {\bibfield  {journal} {\bibinfo  {journal} {Tr. Mosk. Mat. Obs.}\ }\textbf
  {\bibinfo {volume} {4}},\ \bibinfo {pages} {125--236} (\bibinfo {year}
  {1955})}\BibitemShut {NoStop}%
\bibitem [{\citenamefont {Lee}\ \emph {et~al.}(2014)\citenamefont {Lee},
  \citenamefont {Hsieh}, \citenamefont {Flammia},\ and\ \citenamefont
  {Lee}}]{lee14}%
  \BibitemOpen
  \bibfield  {author} {\bibinfo {author} {\bibfnamefont {Yi-Chan}\ \bibnamefont
  {Lee}}, \bibinfo {author} {\bibfnamefont {Min-Hsiu}\ \bibnamefont {Hsieh}},
  \bibinfo {author} {\bibfnamefont {Steven~T.}\ \bibnamefont {Flammia}}, \ and\
  \bibinfo {author} {\bibfnamefont {Ray-Kuang}\ \bibnamefont {Lee}},\
  }\bibfield  {title} {\enquote {\bibinfo {title} {Local {PT} {Symmetry}
  {Violates} the {No}-{Signaling} {Principle}},}\ }\href {\doibase
  10.1103/PhysRevLett.112.130404} {\bibfield  {journal} {\bibinfo  {journal}
  {Physical Review Letters}\ }\textbf {\bibinfo {volume} {112}},\ \bibinfo
  {pages} {130404} (\bibinfo {year} {2014})}\BibitemShut {NoStop}%
\bibitem [{\citenamefont {Zhan}\ \emph {et~al.}(2020)\citenamefont {Zhan},
  \citenamefont {Wang}, \citenamefont {Xiao}, \citenamefont {Bian},
  \citenamefont {Zhang}, \citenamefont {Sanders}, \citenamefont {Zhang},\ and\
  \citenamefont {Xue}}]{zhan20}%
  \BibitemOpen
  \bibfield  {author} {\bibinfo {author} {\bibfnamefont {Xiang}\ \bibnamefont
  {Zhan}}, \bibinfo {author} {\bibfnamefont {Kunkun}\ \bibnamefont {Wang}},
  \bibinfo {author} {\bibfnamefont {Lei}\ \bibnamefont {Xiao}}, \bibinfo
  {author} {\bibfnamefont {Zhihao}\ \bibnamefont {Bian}}, \bibinfo {author}
  {\bibfnamefont {Yongsheng}\ \bibnamefont {Zhang}}, \bibinfo {author}
  {\bibfnamefont {Barry~C.}\ \bibnamefont {Sanders}}, \bibinfo {author}
  {\bibfnamefont {Chengjie}\ \bibnamefont {Zhang}}, \ and\ \bibinfo {author}
  {\bibfnamefont {Peng}\ \bibnamefont {Xue}},\ }\bibfield  {title} {\enquote
  {\bibinfo {title} {Experimental quantum cloning in a pseudo-unitary
  system},}\ }\href {\doibase 10.1103/PhysRevA.101.010302} {\bibfield
  {journal} {\bibinfo  {journal} {Physical Review A}\ }\textbf {\bibinfo
  {volume} {101}},\ \bibinfo {pages} {010302} (\bibinfo {year}
  {2020})}\BibitemShut {NoStop}%
\bibitem [{\citenamefont {Redheffer}(1962)}]{redheffer62}%
  \BibitemOpen
  \bibfield  {author} {\bibinfo {author} {\bibfnamefont {Raymond}\ \bibnamefont
  {Redheffer}},\ }\bibfield  {title} {\enquote {\bibinfo {title} {On the
  {Relation} of {Transmission}-{Line} {Theory} to {Scattering} and
  {Transfer}},}\ }\href {\doibase 10.1002/sapm19624111} {\bibfield  {journal}
  {\bibinfo  {journal} {Journal of Mathematics and Physics}\ }\textbf {\bibinfo
  {volume} {41}},\ \bibinfo {pages} {1--41} (\bibinfo {year}
  {1962})}\BibitemShut {NoStop}%
\bibitem [{\citenamefont {Born}\ and\ \citenamefont {Wolf}(1999)}]{born_wolf}%
  \BibitemOpen
  \bibfield  {author} {\bibinfo {author} {\bibfnamefont {Max}\ \bibnamefont
  {Born}}\ and\ \bibinfo {author} {\bibfnamefont {Emil}\ \bibnamefont {Wolf}},\
  }\href@noop {} {\emph {\bibinfo {title} {Principles of Optics:
  Electromagnetic Theory of Propagation, Interference and Diffraction of
  Light}}}\ (\bibinfo  {publisher} {Cambridge University Press},\ \bibinfo
  {address} {Cambridge},\ \bibinfo {year} {1999})\BibitemShut {NoStop}%
\bibitem [{\citenamefont {Mostafazadeh}(2020)}]{mosta20}%
  \BibitemOpen
  \bibfield  {author} {\bibinfo {author} {\bibfnamefont {Ali}\ \bibnamefont
  {Mostafazadeh}},\ }\href {\doibase 10.48550/arXiv.2009.10507} {\emph
  {\bibinfo {title} {Transfer matrix in scattering theory: {A} survey of basic
  properties and recent developments}}},\ \bibinfo {type} {Tech. Rep.}\
  (\bibinfo {year} {2020})\ \bibinfo {note} {arXiv:2009.10507 [cond-mat,
  physics:hep-th, physics:math-ph, physics:physics, physics:quant-ph] type:
  article}\BibitemShut {NoStop}%
\bibitem [{\citenamefont {Wheeler}\ and\ \citenamefont
  {Feynman}(1949)}]{wheeler49}%
  \BibitemOpen
  \bibfield  {author} {\bibinfo {author} {\bibfnamefont {John~Archibald}\
  \bibnamefont {Wheeler}}\ and\ \bibinfo {author} {\bibfnamefont
  {Richard~Phillips}\ \bibnamefont {Feynman}},\ }\bibfield  {title} {\enquote
  {\bibinfo {title} {Classical {Electrodynamics} in {Terms} of {Direct}
  {Interparticle} {Action}},}\ }\href {\doibase 10.1103/RevModPhys.21.425}
  {\bibfield  {journal} {\bibinfo  {journal} {Reviews of Modern Physics}\
  }\textbf {\bibinfo {volume} {21}},\ \bibinfo {pages} {425--433} (\bibinfo
  {year} {1949})}\BibitemShut {NoStop}%
\bibitem [{\citenamefont {Feynman}(1949)}]{feynman49}%
  \BibitemOpen
  \bibfield  {author} {\bibinfo {author} {\bibfnamefont {R.~P.}\ \bibnamefont
  {Feynman}},\ }\bibfield  {title} {\enquote {\bibinfo {title} {The {Theory} of
  {Positrons}},}\ }\href {\doibase 10.1103/PhysRev.76.749} {\bibfield
  {journal} {\bibinfo  {journal} {Physical Review}\ }\textbf {\bibinfo {volume}
  {76}},\ \bibinfo {pages} {749--759} (\bibinfo {year} {1949})}\BibitemShut
  {NoStop}%
\bibitem [{\citenamefont {Watanabe}(1955)}]{watanabe}%
  \BibitemOpen
  \bibfield  {author} {\bibinfo {author} {\bibfnamefont {Satosi}\ \bibnamefont
  {Watanabe}},\ }\bibfield  {title} {\enquote {\bibinfo {title} {{Symmetry of
  Physical Laws. Part III. Prediction and Retrodiction}},}\ }\href {\doibase
  10.1103/RevModPhys.27.179} {\bibfield  {journal} {\bibinfo  {journal} {Review
  of Modern Physics}\ }\textbf {\bibinfo {volume} {27}},\ \bibinfo {pages}
  {179--186} (\bibinfo {year} {1955})}\BibitemShut {NoStop}%
\bibitem [{\citenamefont {Aharonov}\ and\ \citenamefont
  {Rohrlich}(2005)}]{aharonov_rohrlich}%
  \BibitemOpen
  \bibfield  {author} {\bibinfo {author} {\bibfnamefont {Yakir}\ \bibnamefont
  {Aharonov}}\ and\ \bibinfo {author} {\bibfnamefont {Daniel}\ \bibnamefont
  {Rohrlich}},\ }\href {\doibase 10.1002/9783527619115} {\emph {\bibinfo
  {title} {{Quantum Paradoxes: Quantum Theory for the Perplexed}}}}\ (\bibinfo
  {publisher} {Wiley},\ \bibinfo {address} {New York},\ \bibinfo {year}
  {2005})\BibitemShut {NoStop}%
\bibitem [{\citenamefont {Cramer}(1986)}]{cramer86}%
  \BibitemOpen
  \bibfield  {author} {\bibinfo {author} {\bibfnamefont {John~G.}\ \bibnamefont
  {Cramer}},\ }\bibfield  {title} {\enquote {\bibinfo {title} {The
  transactional interpretation of quantum mechanics},}\ }\href {\doibase
  10.1103/RevModPhys.58.647} {\bibfield  {journal} {\bibinfo  {journal}
  {Reviews of Modern Physics}\ }\textbf {\bibinfo {volume} {58}},\ \bibinfo
  {pages} {647--687} (\bibinfo {year} {1986})}\BibitemShut {NoStop}%
\bibitem [{\citenamefont {Tsang}(2009)}]{smooth}%
  \BibitemOpen
  \bibfield  {author} {\bibinfo {author} {\bibfnamefont {Mankei}\ \bibnamefont
  {Tsang}},\ }\bibfield  {title} {\enquote {\bibinfo {title} {Time-symmetric
  quantum theory of smoothing},}\ }\href {\doibase
  10.1103/PhysRevLett.102.250403} {\bibfield  {journal} {\bibinfo  {journal}
  {Physical Review Letters}\ }\textbf {\bibinfo {volume} {102}},\ \bibinfo
  {pages} {250403} (\bibinfo {year} {2009})}\BibitemShut {NoStop}%
\bibitem [{\citenamefont {Bogaerts}\ \emph {et~al.}(2020)\citenamefont
  {Bogaerts}, \citenamefont {Pérez}, \citenamefont {Capmany}, \citenamefont
  {Miller}, \citenamefont {Poon}, \citenamefont {Englund}, \citenamefont
  {Morichetti},\ and\ \citenamefont {Melloni}}]{bogaerts20}%
  \BibitemOpen
  \bibfield  {author} {\bibinfo {author} {\bibfnamefont {Wim}\ \bibnamefont
  {Bogaerts}}, \bibinfo {author} {\bibfnamefont {Daniel}\ \bibnamefont
  {Pérez}}, \bibinfo {author} {\bibfnamefont {José}\ \bibnamefont {Capmany}},
  \bibinfo {author} {\bibfnamefont {David A.~B.}\ \bibnamefont {Miller}},
  \bibinfo {author} {\bibfnamefont {Joyce}\ \bibnamefont {Poon}}, \bibinfo
  {author} {\bibfnamefont {Dirk}\ \bibnamefont {Englund}}, \bibinfo {author}
  {\bibfnamefont {Francesco}\ \bibnamefont {Morichetti}}, \ and\ \bibinfo
  {author} {\bibfnamefont {Andrea}\ \bibnamefont {Melloni}},\ }\bibfield
  {title} {\enquote {\bibinfo {title} {Programmable photonic circuits},}\
  }\href {\doibase 10.1038/s41586-020-2764-0} {\bibfield  {journal} {\bibinfo
  {journal} {Nature}\ }\textbf {\bibinfo {volume} {586}},\ \bibinfo {pages}
  {207--216} (\bibinfo {year} {2020})}\BibitemShut {NoStop}%
\bibitem [{\citenamefont {Reck}\ \emph {et~al.}(1994)\citenamefont {Reck},
  \citenamefont {Zeilinger}, \citenamefont {Bernstein},\ and\ \citenamefont
  {Bertani}}]{reck94}%
  \BibitemOpen
  \bibfield  {author} {\bibinfo {author} {\bibfnamefont {Michael}\ \bibnamefont
  {Reck}}, \bibinfo {author} {\bibfnamefont {Anton}\ \bibnamefont {Zeilinger}},
  \bibinfo {author} {\bibfnamefont {Herbert~J.}\ \bibnamefont {Bernstein}}, \
  and\ \bibinfo {author} {\bibfnamefont {Philip}\ \bibnamefont {Bertani}},\
  }\bibfield  {title} {\enquote {\bibinfo {title} {Experimental realization of
  any discrete unitary operator},}\ }\href {\doibase 10.1103/PhysRevLett.73.58}
  {\bibfield  {journal} {\bibinfo  {journal} {Physical Review Letters}\
  }\textbf {\bibinfo {volume} {73}},\ \bibinfo {pages} {58--61} (\bibinfo
  {year} {1994})}\BibitemShut {NoStop}%
\end{thebibliography}%

\end{document}